\documentclass{iopart}
\usepackage{epsfig}
\usepackage{latexsym}

\begin{document}

\newcommand{\tbox}[1]{\mbox{\tiny #1}}
\newcommand{\half}{\mbox{\small $\frac{1}{2}$}}
\newcommand{\mbf}[1]{{\mathbf #1}}

\newcommand{\mV}{{\mathsf{V}}}
\newcommand{\mL}{{\mathsf{L}}}
\newcommand{\mA}{{\mathsf{A}}}
\def\ve{v_{\tbox{E}}}			
\newcommand{\lB}{\lambda_{\tbox{B}}}  
\newcommand{\ofr}{{(\mbf{r})}}       
\newcommand{\ofs}{{(\mbf{s})}}       
\def\eg{{\it e.g.\ }}
\def\ie{{\it i.e.\ }}
\def\etal{{\it et al.\ }}

\def\be{\begin{equation}}
\def\ee{\end{equation}}
\def\bea{\begin{eqnarray}}
\def\eea{\end{eqnarray}}

\def\tcl{\tau_{\tbox{cl}}}		
\def\tbl{\tau_{\tbox{bl}}}		
\def\tcol{\tau_{\tbox{col}}}		
\def\cew{\tilde{C}_{\tbox E}(\omega)}		
\def\cw{\tilde{C}(\omega)}		
\def\terg{t_{\tbox{erg}}}		

\newcommand{\rop}{\hat{\mbf{r}}}	
\newcommand{\pop}{\hat{\mbf{p}}}


\title{Rate of energy absorption for a driven chaotic cavity}

\author{Alex Barnett$^a$, Doron Cohen$^a$ 
and Eric J. Heller$^{a,b}$}

\date{June 2000}

\address{
$^a$ Department of Physics, Harvard University; \\ 
$^b$ Department of Chemistry and Chemical Biology,  
Harvard University; \\ Cambridge, MA 02138, USA}


\begin{abstract}
We consider the response of a chaotic cavity in $d$ dimensions
to periodic driving.
We are motivated by older studies of 
one-body dissipation in nuclei, 
and also by anticipated mesoscopic applications.
For calculating the rate of energy absorption 
due to time-dependent deformation of the 
confining potential, we introduce an improved version 
of the wall formula. Our formulation takes into account 
that a special class of deformations causes no heating 
in the zero-frequency limit.
We also derive a mesoscopic version of the Drude formula, 
and explain that it can be regarded as a special example 
of our calculations. 
Specifically we consider a quantum dot driven 
by electro-motive force which is induced by 
a time-dependent homogeneous magnetic field. 
\end{abstract}


\section{Introduction}

The dynamics of a particle inside a cavity
(billiard) in $d=2$ or 3 dimensions 
is major theme in studies of classical and quantum chaos. 
Whereas the physics of time independent chaotic systems 
is extensively explored, less is known 
about the physics of time-dependent chaotic systems. 
The main exceptions are the studies of 
the kicked rotator and related systems \cite{qkr}. 
However, the rotator (with no kicks) 
is a 1D integrable system, whereas we 
are interested in chaotic (2D or 3D) cavities.

Driven cavities were of special interest in the 
studies of one-body dissipation in nuclei 
\cite{wall,koonin,jarz92,jarz93}. 
A renewed interest in this problem is anticipated 
in the field of mesoscopic physics. Quantum dots 
can be regarded as small 2D cavities whose shape 
is controlled by electrical gates. Another variation 
is driving a quantum dot by time-dependent magnetic field. 
In Section~6 we will explain that the calculation of 
the system response in the latter case can be 
regarded as a special example of the study in this paper. 
A similar observation applies to the case of a quantum dot driven 
by a homogeneous time-dependent electric field.
However, in the latter case it is essential to take screening 
into account \cite{grains}, and therefore our calculations no 
longer apply.

We consider a system of non-interacting particles 
inside a cavity whose walls can be deformed.
We define a single parameter $x$ that controls 
this deformation. We would like to consider the 
case where $x(t)=A\sin(\Omega t)$ is being changed  
periodically in time, where $A$ is the amplitude  
and $\Omega$ is the driving frequency. In particular 
we are interested in the small frequency limit, 
meaning $\Omega \ll 1/\tau_{\tbox{col}}$. 
Here $\tau_{\tbox{col}}$ is the typical time between 
collisions with the moving walls of the cavity.

We will be interested in general deformations which need not 
preserve the billiard shape nor its volume.
We can specify any deformation by a function $D(\mbf{s})$,
where~$\mbf{s}$ specifies the location of a wall element 
on the boundary (surface) of the cavity, and $D(\mbf{s})\delta x$ is the 
normal displacement of this wall element.
There is a restricted class of deformations that are shape-preserving:
they involve translations, rotations and dilations of the cavity.
We will see that this class has special properties. Note that 
translations and rotations are also volume-preserving, in
which case
the associated time-dependent deformations
can be described as `shaking' the cavity.

What is the rate in which the `gas' inside the cavity is 
heated up?  The answer depends on the shape 
of the cavity, the deformation $D(\mbf{s})$ involved, 
as well as on the amplitude $A$ and the driving frequency $\Omega$. 
Also the number of particles $N$ and their energy 
distribution $\rho(E)$ should be specified.  

For non-interacting particles the solution of this problem 
is reduced to the analysis of one-particle physics. This 
observation is self-evident for non-interacting classical particles, 
but it is also true for non-interacting fermions (see \ref{ap:lrt}). 
We would like to work within the framework of linear 
response theory (LRT). In such case one can write 
%
\begin{eqnarray} \label{e1a} 
\frac{d}{dt}\langle {\cal H} \rangle  = 
\mu(\Omega) \times \half  (A\Omega)^2
\end{eqnarray}
where the dissipation coefficient $\mu(\Omega)$ 
is amplitude independent. The small-$\Omega$ version 
of this formula can be written as 
\begin{eqnarray} \label{e1b}  
\frac{d}{dt}\langle {\cal H} \rangle  = 
\mu V^2  
\end{eqnarray}
%
where $\mu=\mu(\Omega \rightarrow 0)$ is known as 
the friction coefficient, and $V = A\Omega/\sqrt{2}$ 
is the average root-mean-square (RMS) deformation velocity.  
For convenience let us define $x$ as having  
units of length, such that $V$ characterizes 
the velocity of the moving walls.

A necessary classical condition for the validity of LRT 
is $V \ll \ve$ where $\ve \equiv (2E/m)^{1/2}$ is the velocity 
of the particle \cite{vrn,frc}.
We also assume that the motion of the particle inside 
the cavity is globally chaotic, meaning no mixed phase space 
\cite{mixed}.
The criteria for having such a cavity are discussed in \cite{bunim,design}.
The justification of LRT in the quantum-mechanical case 
is more subtle \cite{frc,crs,rsp}, and does not constitute 
a theme in this paper,
although we do connect with the quantum case in Section~\ref{quant}.
The theory to be presented 
assumes that LRT is a valid formulation of the problem.

As explained in \ref{ap:lrt}, LRT implies that the 
dissipation coefficient $\mu(\Omega)$ is related 
via a fluctuation-dissipation (FD) relation to a spectral function
$\tilde{C}_{\tbox{E}}(\Omega)$. Namely,  
\begin{eqnarray} \label{e2}  
\mu(\Omega) \ = 
\ \frac{1}{2} \int_0^{\infty} \rho(E) dE 
\ \frac{1}{g(E)}
\frac{\partial}{\partial E}
\left[g(E) \tilde{C}_{\tbox{E}}(\Omega) \right]
\end{eqnarray}
Here $\rho(E)$ is the energy distribution of the 
particles, and $g(E)$ is the density of states. 
The function $\cew$ is the noise power spectrum of 
the generalized `force' associated with the parameter $x$. 
This function is the main object of the present study, 
and its precise definition is in Section~\ref{sec:sys}.
We shall chiefly explore how $\cew$ depends on the type of 
deformation involved, but also discuss effects due to the cavity shape.

In particular we are interested in the small frequency 
limit where $\mu$ is related to the fluctuations intensity 
\begin{eqnarray} \label{e3}  
	\nu_{\tbox{E}} \ = \ \tilde{C}_{\tbox{E}}(0) \ = \ 
	\int_{-\infty}^{\infty} \! C_{\tbox{E}}(\tau) d\tau .
\end{eqnarray}
The simplest estimate for $\nu_{\tbox{E}}$, which we are going to
call `white noise approximation' (WNA), leads 
(in case of a 3D cavity) to the well known `wall formula' \cite{wall}
\begin{eqnarray} \label{e4}  
	\mu_{\tbox{E}} = \frac{N}{\mathsf{V}} 
	m \ve \oint D(\mbf{s})^2 d\mbf{s}
\end{eqnarray}
where the subscript $\tbox{E}$ implies that we are considering 
a microcanonical ensemble $\rho(E)$, the number 
of particles is $N$, and the volume of the cavity 
is $\mV$.
The above version of the wall formula has been derived 
for the purpose of calculating the so-called one-body 
dissipation rate in nuclei.
The original 
derivation of this formula is based on a simplified
kinetic picture \cite{wall}. For alternate derivations using the
LRT approach see \cite{koonin}. For the generalization to 
any dimension $d$ using the LRT-FD strategy see \cite{frc}
and further references therein.

Our main purpose is to introduce an improved version of the 
wall formula, and to analyze the frequency dependence 
of $\mu(\Omega)$.    
This will involve a demonstration \cite{dil} that for special types
of deformations (namely dilations, translations and rotations) 
the small-$\Omega$ dissipation rate is remarkably 
different from the naive expectation.
As an application, the mesoscopic version of Drude formula for the 
{\em conductance} of a quantum dot
in a uniform
time-dependent magnetic field
reduces to the 
the calculation of $\cew$ for one of these
special deformations (namely rotation), and leads to 
(see Section~\ref{sec:drude}) 
\begin{eqnarray} \label{e5} 
\mu(\Omega) \ \sim \ \frac{N}{{\mathsf A}}
\left(\frac{e^2}{m}\tau_{\tbox{col}}\right) 
\frac{1}{1+(\tau_{\tbox{col}} \Omega)^2}  
\end{eqnarray} 
where ${\mathsf A}$ is the area of the dot.

For our improved wall formula,
we show that it is essential to 
project out the special components of a general deformation, 
and only then to estimate $\nu_{\tbox{E}}$ using the WNA.
If the assumption of strong chaos cannot be justified, 
further corrections are required due to correlations 
between successive bounces.

The effect of interaction between the particles is 
not discussed in this paper. If the mean free path 
for inter-particle collisions is large compared 
with the size of the cavity, then we expect that our 
analysis still applies. If the mean free path is 
much smaller, then we get into the hydrodynamic regime. 
In the latter case we have a drag effect, and the dissipation 
rate is determined by the viscosity of the gas 
via Stokes' law.

\section{The model system}
\label{sec:sys}

Consider a particle whose canonical coordinates 
are $(\mbf{r},\mbf{p})$ moving inside a cavity. 
The Hamiltonian is    
\begin{eqnarray} \label{e6} 
{\cal H}(\mbf{r},\mbf{p};x) = 
\mbf{p}^2/2m + U( \mbf{r} - x\mbf{D}(\mbf{r}) ) 
\end{eqnarray} 
where $U(\mbf{r})$ is the confining potential. 
We have introduced a (unitless) deformation `field'
$\mbf{D}(\mbf{r})$, and $x$ is the controlling 
parameter. In this paper we assume that 
$U(\mbf{r})=0$ inside the cavity. The volume of the 
cavity is ${\mathsf V}$. Outside the cavity the 
potential $U(\mbf{r})$ becomes very large. 
To be specific, one may assume that the walls 
exert a normal force $f$, and we take the
hard wall limit $f \rightarrow \infty$. 
With the above assumptions about $U(\mbf{r})$ it is 
clear that the deformation is completely specified 
by the boundary function
$D(\mbf{s}) \equiv \hat{\mbf{n}}(\mbf{s})\cdot\mbf{D}(\mbf{s})$, 
where $\hat{\mbf{n}}(\mbf{s})$ is an outwards 
unit normal vector at the boundary point $\mbf{s}$.

\begin{figure}
\centerline{\epsfig{figure=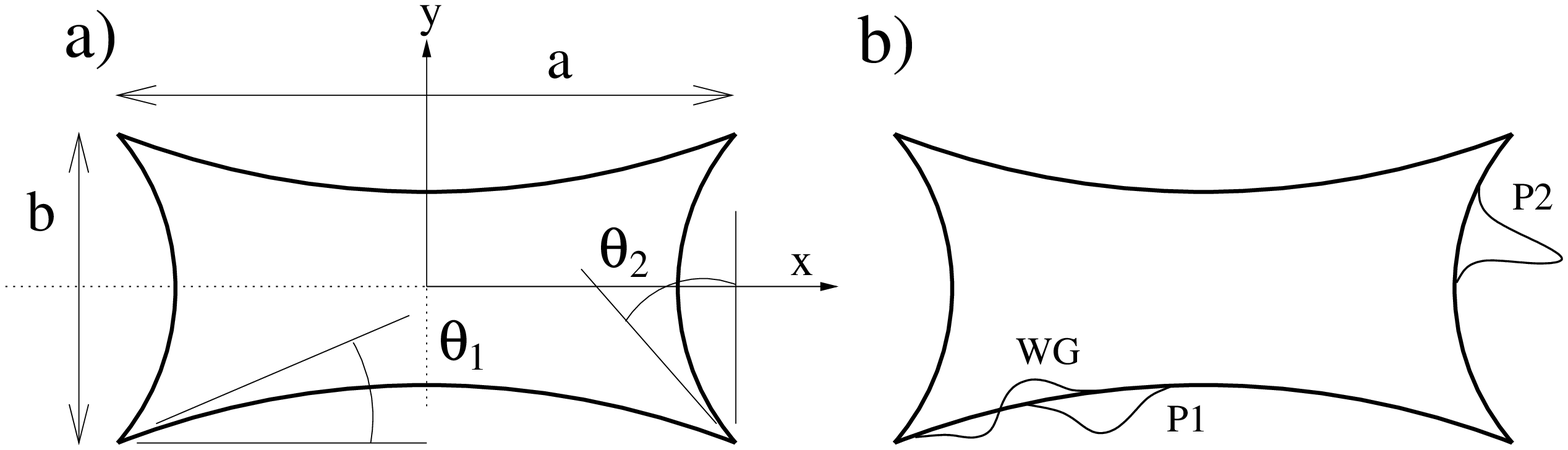,width=0.7\hsize}}
\caption{
{\bf a}) The generalized two-dimensional Sinai billiard 
which has been used for our numerical studies.  
{\bf b})~Three example deformations are illustrated.  
Note that they are shown exaggerated in strength.
}
\label{fig:gsinai}
\end{figure}

Most of our numerical tests will refer to the 2D cavity
illustrated in Fig.~\ref{fig:gsinai}a.
It is a generalized two-dimensional Sinai billiard
formed from concave arcs of circles with two different radii.
Typical parameters used are $a{=}2$, $b{=}1$,
$\theta_1{=}0.2$, $\theta_2{=}0.5$,
for which the average collision rate with the wall is
$(1/\tbl) \approx 0.63$.
This billiard has been chosen because it has `hard chaos': 
There is no mixed phase space, and there are no marginally-stable 
orbits (see Section~\ref{sec:wnarev}).
In Fig.~\ref{fig:gsinai}b we show three example deformations.
For illustration purposes we have selected three `localized' 
deformations.
See Tables~1 and 2 for a full list of deformations that have
been tested in our numerical work.

\begin{figure}
\centerline{\epsfig{figure=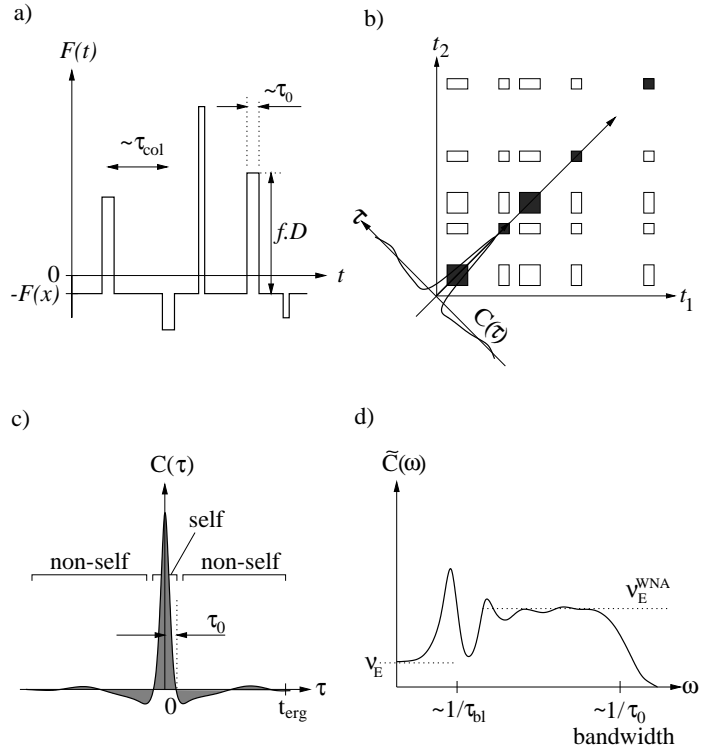,width=0.7\hsize}}
\caption{
The fluctuating force ${\cal F}(t)$ looks 
like a train of impulses ({\bf a}).
Due to ergodicity the autocorrelation 
function $C(\tau)$ can be regarded as a time 
average ({\bf b}).  
The resultant autocorrelation function ({\bf c}) 
and the associated power spectrum ({\bf d}) 
may be characterized by non universal features.
See text for further explanations. 
}
\label{fig:f+c}
\end{figure}

Associated with the parameter $x$ is the fluctuating quantity
\begin{eqnarray} \label{e7} 
{\cal F}(t) =  - \left. 
\frac{\partial {\cal H}(\mbf{r},\mbf{p};x)}{\partial x} 
\right|_{x=0} 
\end{eqnarray}
where the time-dependence arises from that of the trajectory 
$(\mbf{r}(t),\mbf{p}(t))$.
This quantity can be thought of as the generalized time-dependent
`force' associated with the parameter $x$.
For the Hamiltonian (\ref{e6}) we can write
\begin{eqnarray} \label{e9} 
{\cal F}(t) =  \mbf{D}\ofr \cdot {\mbf{\nabla}} U\ofr =  
- \mbf{D}\ofr \cdot \dot{\mbf{p}} .
\end{eqnarray} 
Recognizing $\dot{\mbf{p}}$ as the force on the gas particle,
we see that ${\cal F}(t)$ is a train of spikes (see Fig. \ref{fig:f+c}a).
Namely, the fluctuating force ${\cal F}(t)$ 
consists of impulses whose maximum duration 
is $\tau_0 = 2m \ve /f$.  In the hard wall limit 
$\tau_0 \rightarrow 0$, and we can write 
\begin{eqnarray} \label{e10} 
{\cal F}(t) \ = \   
\sum_{i} 2m \ve \ \cos(\theta_i) 
\ D_i \ \delta(t-t_i) 
\end{eqnarray} 
where $i$ labels collisions: $t_i$ is the time of a collision, 
$D_i$ stands for $D(\mbf{s}_i)$
at the location $\mbf{s}_i$ of a collision, and 
$\ve \cos(\theta_i)$ 
is the normal component of the particle's collision velocity. 
The above sequence of impulses is characterized by 
an average rate of collisions $1/\tau_{\tbox{col}}$. 
The quantitative definition of $\tau_{\tbox{col}}$
is postpone to Section~\ref{sec:wna}. 
Note however that $\tau_{\tbox{col}}$ may be much 
larger that the ballistic time $\tau_{\tbox{bl}}$. 
The ballistic time is the average time 
between collisions with the boundary. 
We have $\tau_{\tbox{col}}\gg\tau_{\tbox{bl}}$
whenever a deformation involves only a small piece 
of the boundary. 
Finally we note that if the deformation is volume-preserving 
then $\langle {\cal F}(t)\rangle =0$.  Otherwise it is 
convenient to subtract the (constant) average value $F(x)$ 
from the above definition of ${\cal F}(t)$. 
This convention is reflected in our illustration (Fig.\ref{fig:f+c}a).

We define the auto-correlation function 
of ${\cal F}(t)$ as follows:
\begin{eqnarray} \label{e8}  
C_{\tbox{E}}(\tau) \; \equiv \;
\langle {\cal F}(t) {\cal F}(t+\tau) \rangle_{\tbox{E}} 
\end{eqnarray} 
The subscript $\tbox E$, whenever used, suggests that the 
average over initial conditions is of microcanonical 
type, with energy $E$.
Note that $C_{\tbox{E}}(\tau)$ is defined using the time independent
(`frozen') Hamiltonian, and therefore is independent of $t$. 
The auto-correlation function $C_{\tbox{E}}(\tau)$
can be handled as a time-average rather than
an ensemble-average (by ergodicity). 
The resulting construction is illustrated in Fig.~\ref{fig:f+c}b, 
where we illustrate the projection of ${\cal F}(t_1){\cal F}(t_2)$
onto the $\tau \equiv t_2-t_1$ axis. 
The contribution for the self correlation is shaded.
The forms of the resultant $C_{\tbox{E}}(\tau)$ 
and its Fourier transform 
$\cew \equiv \int C_{\tbox{E}}(\tau) \exp(i \omega \tau) d\tau$
are illustrated schematically
in Figs.~\ref{fig:f+c}c and \ref{fig:f+c}d.
Note that the $\omega\rightarrow0$ limit of $\cw$
is equal to the area under $C(\tau)$.

The auto-correlation function $C_{\tbox{E}}(\tau)$ 
consists of a $\tau=0$ (`self') peak due to the self-correlation 
of the spikes, and of an additional smooth (`non-self') component 
due to correlations between successive bounces. 
This implies \cite{bouncenote}
that pronounced correlations are usually characterized 
by the time scale $\tau_{\tbox{bl}}$, rather than $\tau_{\tbox{col}}$. 
Consequently the associated frequency scale for non-universal  
structures is $\omega \sim 1/\tau_{\tbox{bl}}$.  Another 
relevant time scale is the ergodic time $\terg$ which is the 
inverse of the average Lyapunov (instability) exponent. 
Beyond $\terg$  the correlations become vanishingly small. 
Non-negligible tails may arise only if the motion has marginally 
stable orbits.

As explained in the Introduction, we shall be most interested 
in the noise intensity $\nu_{\tbox E}$ defined by (\ref{e3}).
Observing that ${\cal F}(t)$ is linear in $D\ofs$,
it follows that the noise intensity is a quadratic functional
\be
\label{eq:kernel}
	\nu_{\tbox E} \; = \; \oint \! \! \oint d\mbf{s}_1 d\mbf{s}_2
	\, D(\mbf{s}_1) \gamma_{\tbox E}(\mbf{s}_1,\mbf{s}_2) D(\mbf{s}_2) ,
\ee
where the kernel $\gamma_{\tbox E}$ depends on both the 
cavity shape and the particle energy $E$ \cite{koonin}.
Furthermore, billiards
are {\em scaling systems} in the sense that a change in $E$ leaves the
trajectories unchanged.
From this and (\ref{e10}) we have the scaling relation
$\gamma_{\tbox E}(\mbf{s}_1,\mbf{s}_2) =
m^2 \ve^3 \cdot \hat{\gamma}(\mbf{s}_1,\mbf{s}_2)$,
where the scaled kernel depends entirely on the geometrical
shape of the cavity.
However, the reason for being interested in {\em approximations} for
$\nu_{\tbox E}$ is that the exact result for the kernel $\hat{\gamma}$
is very complicated to evaluate, and
involves a sum over all classical paths from $\mbf{s}_1$ to $\mbf{s}_2$
(see \cite{koonin}).

\section{Quantum-classical correspondence}
\label{quant}

This paper applies classical physics in order to 
analyze the response of a wide class of systems,
including mesoscopic systems where quantum mechanics may play a role. 
How much of a compromise is a classical 
analysis of the dissipation? This question has been addressed 
in \cite{crs,frc}.  In the level of one-particle physics 
the answer is as follows: within the framework of LRT 
the only difference between the classical formulation 
and the quantal one is involved in replacing 
the classical definition of $C_{\tbox{E}}(\tau)$ by the corresponding 
quantum-mechanical definition. In the level of many 
(non-interacting) particles the only further modification 
is associated with the application of the FD relation, 
as discussed in Appendix A (See Eq.(\ref{e29})). 
We would like to re-emphasize that we assume in this paper 
that we are in an $(A,\Omega)$ regime where LRT is a valid
formulation. The quantum adiabatic regime (extremely small $\Omega$), 
and the non-perturbative regime (see discussion
in \cite{rsp}) are excluded from our considerations.

\begin{figure}
\centerline{\epsfig{figure=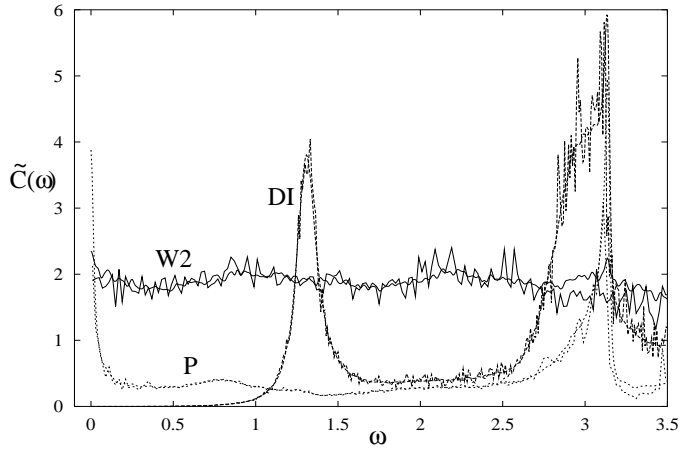,width=0.7\hsize}}
\caption{
Agreement between quantum and classical $\cew$ in the two-dimensional
quarter-stadium billiard for three example deformations (see text).
In each case classical is shown as a thick line,
and quantum a thin line. The $y$-axis has been displaced 
to clearly show the $\omega\rightarrow0$ behavior.
The singular peak at $\omega = \pi$ is due to the `bouncing ball' orbit.
}
\label{fig:qcc}
\end{figure}

Thus the only remaining question is whether 
a classical calculation of $\tilde{C}_{\tbox{E}}(\omega)$ 
is a good approximation quantum-mechanically. 
The answer is that the quantum-classical correspondence 
here is remarkable. It has been tested for a few 
example systems \cite{dil,lds,prm}. In Fig.~\ref{fig:qcc} we demonstrate 
correspondence for the stadium billiard 
for three types of deformations: 
DI (dilation), W2 (periodic oscillation around the perimeter),
and P (wide `piston' existing only on the top edge).
The RMS estimation error is 3\% for the classical calculation 
and 10\% for the quantum calculation.   
The quantum estimate of $\cew$ amounts to computing boundary
overlap integrals of the eigenfunctions (see~\cite{dil}).
We have used all 451 states lying between wavenumbers $398<k<402$,
where the mean level spacing is
$\Delta \approx 8.8 \times 10^{-3}$ in $\omega$ units.
Note that there are $\sim 10^2$ de Broglie
wavelengths across the system.
The stadium was chosen because it enables efficient quantization
using the method of Vergini and Saraceno~\cite{v+s,mythesis}.
An especially good basis set is known for this shape 
\cite{verginithesis}.

\section{The white noise approximation}
\label{sec:wna}

The most naive estimate of the fluctuations intensity 
is based on the WNA. Namely, one assumes that the 
correlation between bounces can be neglected.
This corresponds \cite{koonin} to the {\em local} part of the
kernel (\ref{eq:kernel}). In such case 
only the self-correlation of the spikes is taken into 
consideration and one obtains \cite{frc}
\begin{eqnarray} \label{e11}  
\nu_{\tbox{E}} \ \approx \ 
(2mv_{\tbox{E}})^2 \left\langle 
\sum_{i} \cos^2(\theta_i) 
\ D_i^2 \ \delta(t-t_i) \right\rangle_{\tbox E}
\end{eqnarray}
and from here (see \cite{frc}) using ergodicity,
\begin{eqnarray} \label{e12}  
\label{eq:wna}
\nu_{\tbox{E}} \ \approx \ 
2m^2 v_{\tbox{E}}^3\langle|\cos\theta|^3\rangle 
\frac{1}{{\mathsf V}} \oint [D(\mbf{s})]^2 d\mbf{s}
\end{eqnarray}
where the geometric factor for $d=2,3,\cdots$ is 
$\langle|\cos(\theta)|^3\rangle = 4/(3\pi),1/4,\cdots$.
If we can use the convention $|D(\mbf{s})| \sim 1$ 
over the deformed region (and zero otherwise), 
then  we can write the WNA as 
$\nu_{\tbox{E}} = (2mv_{\tbox{E}})^2 \times (1/\tau_{\tbox{col}})$ 
where $(1/\tau_{\tbox{col}})$ defines the effective
collision rate. For a more careful discussion 
see Appendix F of Ref.\cite{frc}. 
Note again that $\tau_{\tbox{col}}$ can be much larger than the 
ballistic time $\tbl$ in the case that only a small piece 
of the boundary is being deformed.

\begin{figure}
\centerline{\epsfig{figure=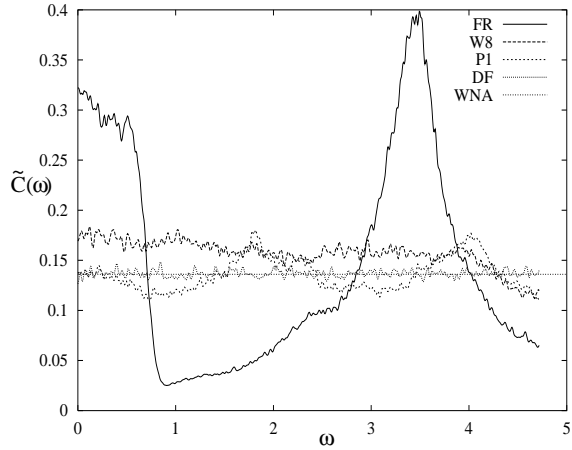,width=0.6\hsize}}
\caption{
The white-noise approximation estimate (WNA is the horizontal dotted line)
compared to actual $\cew$ power spectra
for some example deformations of the 2D generalized Sinai
billiard, with $m = v = 1$.
(The RMS estimation
error of 3\% can be seen as
multiplicative noise with short correlation length in $\omega$).
Deformation functions are defined in Table~I,
and discussed further in the text.
}
\label{fig:wna_corr}
\end{figure}

The use of the WNA can be justified whenever  
successive collisions are effectively 
uncorrelated.  The applicability of such  an
assumption depends on the shape of the cavity
(which will determine the decay of correlations via the
typical Lyapunov exponent)
as well as on the type of deformation involved. 
If we have the cavity of Fig.~\ref{fig:gsinai}a, and the deformation 
involves only a small piece of the boundary (\eg see Fig.~\ref{fig:gsinai}b), 
then successive collisions with the {\em deformed part} of the boundary 
are effectively uncorrelated.
This is so because there 
are many collisions with static pieces of the boundary 
before the next effective collision (with non zero $D_i$)
takes place.
If the deformation involves a large 
piece (or all) of the boundary, we can still argue 
that successive collisions are effectively uncorrelated
provided  $D(\mbf{s})$ is `oscillatory' enough
(ie changes sign many times along the boundary).
These expectations are qualitatively confirmed 
by the numerical results of Fig.~\ref{fig:wna_corr}.
Here
we show a sequence of deformation types
for which the WNA performs increasingly well:
FR (for which sensitivity to the vertical least-unstable periodic orbit
causes large correlation effects and large deviations from WNA),
W8 (oscillatory deformation changes sign many times around the perimeter,
giving better agreement with WNA),
P1 (localized `piston' type deformation, for which WNA is good),
and
DF (random function of zero correlation-length along the
perimeter, showing complete WNA agreement).

The numerical evaluation of $\cew$ throughout this work
is performed by squaring the Fourier transform of a single long
sample of ${\cal F}(t)$ ( $\sim 10^6$ consecutive collisions).
Ergodicity ensures that the properties of a single trajectory
reproduce the desired ensemble-average $\langle \cdots \rangle_{\tbox E}$.
In practice the power spectrum of a single sample is a
stochastic quantity with no correlations in $\omega$-space.
To estimate the underlying noise spectrum $\cew$ 
a smoothing convolution in $\omega$-space is performed.
In the figures a smoothing width of $10^{-2}$ is typical,
giving 3\% RMS estimation error.
The $\delta$-function nature of ${\cal F}(t)$ is handled 
by convolving in the time-domain with a suitably-narrow Gaussian. 
This enables the signal to be sampled uniformly in time,
and hence we can benefit from use of
the Fast Fourier Transform procedure.

It might be asked whether the exponential growth in
sensitivity to numerical round-off error invalidates the
computation of the properties of a long
classical trajectory.
The answer is no: it has been shown that in simple two-dimensional
chaotic maps such as ours, a numerically-generated `pseudo-trajectory'
{\em shadows} (is very close to) a true trajectory with slightly
different initial conditions \cite{shadowing}.
However, as we shall see, the differences in $\omega\rightarrow0$
behavior (in the hard chaos case)
do not in fact rely on correlation properties over times
any longer than $\terg$.

\begin{table}
\label{tbl:deftypes}
\centerline{
\begin{tabular}{lll}
{\em key} & {\em description} & {\em surface deformation function $D(s)$}
	\\ \hline
CO   & constant    & 1 \\
W$n$ & $n$ periods & $\cos(2 \pi n s/L)$ \\
DF   & diffuse     & random[-1,1] (equivalent to W$\infty$) \\
FR   & fracture    & sgn$(x(s))$ if on top or bottom, else 0 \\
SX   & shift-x     & sgn$(x(s))$ if on left or right, else 0 \\
P1   & piston 1    & $10 \exp(-\half \alpha^2)$,
	$\ \alpha{=}(s/L - 0.3)/0.01$ \\
P2   & piston 2    & $10 \exp(-\half \alpha^2)$,
	$\ \alpha{=}(s/L - 0.6)/0.005$ \\
WG   & wiggle      & $5\alpha \exp(-\half \alpha^2)$, 
	$\ \alpha{=}(s/L - 0.25)/0.02$ \\
\end{tabular}
}
\caption{
Key to deformation types used for numerical 2D billiard
experiments in this paper. $L$ is the billiard perimeter.
The deformation is described by a function $D(s)$, 
where $s$ is measured counter-clockwise along the 
perimeter with $s=0$ at the upper left corner. In the 
`fracture' and `shift-x' cases we use the horizontal 
Cartesian coordinate $x(s)$.    
}
\end{table}

\section{`Special' deformations}
\label{sec:spec}

\begin{figure}
\centerline{\epsfig{figure=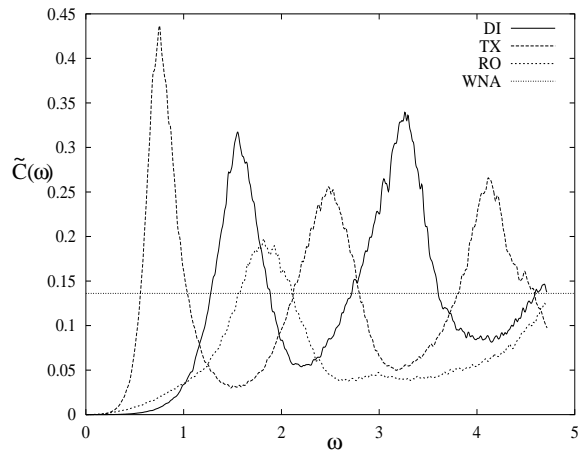,width=0.6\hsize}}
\caption{
The WNA estimate compared to actual $\cw$ 
noise power spectra for example `special' deformation types: 
DI (dilation), TX (translation) and RO (rotation).
See Table~II for definitions.
The WNA fails to predict the vanishing in the small $\omega$ limit.
}
\label{fig:wna_spec}
\end{figure}

The WNA dramatically fails (see Fig.~\ref{fig:wna_spec}) 
for dilation, translations and rotations
(see Table~\ref{tbl:specdefs} for their definitions in 2D).
It is not surprising  that the WNA is `bad' for these  
deformations because their $D(\mbf{s})$ are slowly-changing 
delocalized functions of $s$. However, what is remarkable 
is that $\cew$ for this type of deformations {\em vanishes} 
in the limit $\omega\rightarrow 0$.
Such deformations we would like to call `special' \cite{dil}.
More generally, we would like to say that 
a deformation is `special' if the associated 
fluctuation intensity is $\nu_{\tbox{E}}=0$.

A result that follows from 
the considerations of \ref{ap:alex} 
is that a linear combination of special 
deformation is also special. Therefore the 
special deformations constitute a linear space 
of functions.  We believe that this linear space 
is spanned by the following basis functions: 
one dilation, $d$ translations, and $d(d-1)/2$ rotations. 
However we are not able to give a rigorous mathematical 
argument that excludes the possibility of having 
a larger linear space. In other words, we believe 
that any special deformation can be written as 
a linear combination of dilation, translations 
and rotations.

%

\begin{table}
\label{tbl:specdefs}
\centerline{\begin{tabular}{lll}
{\em key} & {\em description} & {\em deformation field} \\ 
\hline
DI & dilation about origin & $\mbf{D}(\mbf{r}) = \mbf{r}$ \\
TX & $x$-translation & $\mbf{D}(\mbf{r}) = \mbf{e}_x$ \\
TY & $y$-translation & $\mbf{D}(\mbf{r}) = \mbf{e}_y$ \\
RO & rotation about origin &
	$\mbf{D}(\mbf{r}) = \mbf{e}_z \times \mbf{r}$ \\
\end{tabular}
}
\caption{
Key to the four `special' deformations in 2D.
The unit vectors $\mbf{e}_x$ and $\mbf{e}_y$ are 
in the plane (see Fig.~1),
and $\mbf{e}_z$ is in the 
perpendicular direction.
In the case of dilation and rotation $\mbf{D}$
could be made unitless by dividing by a constant length.
}
\end{table}

We will explain the observed $\nu_{\tbox{E}}=0$, 
starting with the case of translations and dilations. 
For translations we have $\mbf{D}=\mbf{e}$, 
where $\mbf{e}$ is a constant vector  
that defines a direction in space. We can write 
${\cal F}(t)=(d/dt)^2 {\cal G}(t)$ 
where ${\cal G}(t)=-m\mbf{e}\cdot\mbf{r}$. 
A similar relation holds for dilation $\mbf{D}=\mbf{r}$ with 
${\cal G}(t)=-\half m \mbf{r}^2$. It follows that 
$\tilde{C}(\omega)=\omega^4\tilde{C}_G(\omega)$, 
where $\tilde{C}_G(\omega)$ is the power spectrum 
of ${\cal G}(t)$.
If $\tilde{C}_G(\omega)$ is a bounded function 
(as it must be when correlations are short-range),
it immediately follows that $\tilde{C}(0) = 0$.
Moreover since ${\cal G}(t)$ is a simple
function of the particle position, we can assume it is a
fluctuating quantity that looks like white 
noise on timescales $> \terg$.
It follows that $\tilde{C}(\omega)$ is 
generically characterized by $\omega^4$ behavior 
for either translations or dilations.

We now turn to consider the case of rotations. This 
case is of particular interest because of its relation 
to the Drude conductance calculation in a
uniform driving magnetic field
(see the following section). 
For rotations we have $\mbf{D}=\mbf{e}\times\mbf{r}$, 
and we can write  ${\cal F}(t)=(d/dt) {\cal G}(t)$, 
where ${\cal G}(t)=-\mbf{e}\cdot(\mbf{r}\times\mbf{p})$,
is a projection of the particle's angular momentum vector
\cite{crossprod}.
Consequently $\tilde{C}(\omega) = \omega^2\tilde{C}_G(\omega)$.
Assuming the angular momentum is a
fluctuating quantity that looks like white 
noise on timescales $> \terg$,
it follows that $\tilde{C}(\omega)$ is
generically characterized by $\omega^2$ behavior.

Thus we have predictions for the power-laws in
the regime $\omega < 1/\terg$ for special deformations
(assuming hard chaos).
These have been verified numerically in our previous 
work~\cite{dil}, with a special emphasis on the case 
of dilation. The case of dilation plays a vital role 
in a highly-successful numerical billiard 
diagonalization method that has been introduced 
recently~\cite{v+s}.

For special deformations we have $\tilde{C}(\omega)=0$
in the limit $\omega = 0$, and consequently the dissipation 
coefficient vanishes ($\mu=0$).
It should be noted that for the case of
a general combination of translations and rotations
this result follows from a simpler argument. 
Taking $\Omega\rightarrow0$ 
while keeping $A\Omega$ constant corresponds to
constant deformation velocity ($\dot{x} = $const).
Transforming the time-dependent Hamiltonian into the reference
frame of the cavity (which is uniformly translating
or rotating with constant velocity)
gives a {\em time-independent} Hamiltonian.
In the new reference frame the energy is a constant 
of the motion, which implies that the system 
cannot absorb energy (no dissipation effect),  
and hence we must indeed have $\mu=0$.

\section{Drude mesoscopic conductance for 2D dot}
\label{sec:drude}

\begin{figure}
\centerline{\epsfig{figure=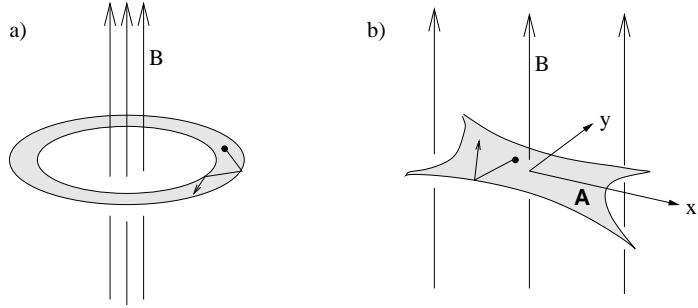,width=0.7\hsize}}
\caption{
Two possible mesoscopic geometries which exhibit
conductance when
driven by a magnetic field:
a) conventional ring of perimeter $\mathsf L$ enclosing the time-dependent flux,
b) ballistic two-dimensional chaotic dot (cavity)
of area $\mathsf{A}$ in a uniform time-dependent
magnetic field.
}
\label{fig:drudegeom}
\end{figure}

Consider a 2D quantum dot in a homogeneous (perpendicular) 
magnetic field (see Fig.~\ref{fig:drudegeom}b).
The one-particle Hamiltonian is 
\begin{eqnarray} \label{e31}
{\cal H}(\mbf{r},\mbf{p};\Phi(t)) \ \ = \ \ 
\frac{1}{2m} 
\left[\mbf{p}-e\mbf{A}(\mbf{r};\Phi(t))\right]^2 
+ U(\mbf{r}) 
\end{eqnarray} 
The dot is defined by the confining potential 
$U(\mbf{r})$, and we choose the magnetic field
as the controlling (driving) parameter.
Periodic driving means $\Phi(t)=A\sin(\Omega t)$. 
The vector potential is given by 
\begin{eqnarray} \label{e32}  
\mbf{A}(\mbf{r};\Phi) = 
\frac{1}{2}\left(\frac{\Phi}{{\mathsf A}} 
\hat{\mbf{z}} \right) \times \mbf{r} 
\end{eqnarray} 
where ${\mathsf A}$ is the area of the dot, 
${\Phi}/{\mathsf A}$ is the magnetic field, 
and $\hat{\mbf{z}}$ is its (perpendicular) direction.

Referring to Eq.(2) one should realize that 
by Faraday's law $V=\dot{\Phi}$ is the 
induced electromotive force (measured in volts). 
Hence $\mu$ is just the conductance.  
The fluctuating quantity that is associated 
with $\Phi$ has the meaning of electric current:
\begin{eqnarray} \label{e33} 
{\cal I}(t) \ = \ 
-\frac{\partial {\cal H}}{\partial \Phi} \ = \ 
\frac{e}{2{\mathsf A}} (\hat{\mbf{z}} \times \mbf{r}) \cdot \mbf{v} 
\end{eqnarray} 
In the conventional ring geometry (Fig.~\ref{fig:drudegeom}a)
the current is just 
${\cal I}(t) = (e/{\mathsf L})v$,
where ${\mathsf L}$ is the perimeter, 
and $v$ is the tangential velocity.
In the general cavity case (Fig.~\ref{fig:drudegeom}b)
${\cal I}(t)$ can be thought of as the angular momentum of the charge.

The Drude mesoscopic conductance is given by the frequency-dependent 
version of the FD relation Eq.(\ref{e29}). With one-particle density 
of states corresponding to 2D gas Eq.(\ref{e29}) becomes   
\begin{eqnarray} \label{e13} 
\mu(\Omega) = \frac{N}{mv_{\tbox{F}}^2}\tilde{C}_{{\cal I}}(\Omega) 
\end{eqnarray}
where the Fermi velocity is related to the Fermi energy  
$E_{\tbox{F}}=\half mv_{\tbox{F}}^2$.  
The power spectrum of the electric current $\tilde{C}_{{\cal I}}(\omega)$ 
is the Fourier transform of the current-current correlation function. 
In standard derivations of the Drude formula it is assumed 
that this correlation function is exponential:
\begin{eqnarray} \label{e14} 
C_{{\cal I}}(\tau) \ \ \sim \ \ 
\frac{e^2}{{\mathsf A}} v_{\tbox{F}}^2
\ \exp\left(-\frac{|\tau|}{\tau_{\tbox{col}}}\right)
\end{eqnarray}
leading to the Lorentzian Eq.(\ref{e5}). 
However, for a given dot shape $C_{{\cal I}}(\tau)$ is not 
really an exponential, but rather reflects the  
system-specific geometry. 
Below we discuss two limits in which we can obtain 
approximations for $C_{{\cal I}}(\tau)$ 
and hence (via Eq.(\ref{e13})) for the 
frequency-dependent conductance $\mu(\Omega)$.

The current ${\cal I}(t)$ is a piecewise constant 
function of time. It is constant between collisions 
with the walls because of conservation of angular momentum. 
The derivative of this quantity, 
${\cal F}(t) = \dot{\cal I}$, is a train of spikes. 
It formally coincides (using (\ref{e9}))
with the ${\cal F}(t)$ of the deformation 
$\mbf{D}(\mbf{r})=(e/(2m{\mathsf A}))\hat{\mbf{z}}\times\mbf{r}$, 
corresponding to rotation around the $z$ axis.
It follows that the current-current correlation $C_{\cal I}(\tau)$ 
is trivially related to the ${\cal F}(t)$ correlation function 
$C(\tau)$ as follows:  
\begin{eqnarray} \label{e34} 
\tilde{C}_{\cal I}(\omega) \ = \ 
\frac{1}{\omega^2} \tilde{C}(\omega) 
\end{eqnarray} 
Thus we see that the calculation of `conductance'  
is formally equivalent to a special case of deformation, 
namely a rotation.

There are two limits in which we can get approximation 
for $\tilde{C}_{\cal I}(\omega)$. For small frequencies  
$\omega\ll(1/\tau_{\tbox{col}})$ we may use the 
following simple estimate:  
\begin{eqnarray} \label{e15} 
\tilde{C}_{{\cal I}}(\omega) \ \ = \ \ 
\langle {\cal I}^2 \rangle \times 2\tau_{\tbox{col}} \ \ = \ \ 
\frac{1}{4} e^2 \ \frac{\langle\mbf{r}^2\rangle}{{\mathsf A}^2}
\ v_{\tbox{F}}^2 \times \tau_{\tbox{col}}
\end{eqnarray}
The first equality can be taken as an operative 
definition of the correlation time $\tau_{\tbox{col}}$ 
in the context of this calculation. 
Obviously, up to a system-specific geometrical 
factor this result ($\sim \omega^0$)
agrees with the standard Drude result.
For ring geometry one should make the replacements 
$\langle\mbf{r}^2\rangle \mapsto ({\mathsf L}/(2\pi))^2$ 
and ${\mathsf A} \mapsto \pi({\mathsf L}/(2\pi))^2$ 
where ${\mathsf L}$ is the length of the wire 
(perimeter of the ring).  Thus one obtains 
$\tilde{C}_{{\cal I}}(\omega) = ((e/{\mathsf L})v_{\tbox{F}})^2$ 
leading to the standard-looking Drude formula for a mesoscopic wire 
$\mu= (N/{\mathsf L}^2)\times(e^2/m)\times\tau_{\tbox{col}}$.

In the limit $\omega\gg (1/\tau_{\tbox{col}})$ we can get 
a much more satisfying result. The fluctuating quantity 
${\cal F}(t)=\dot{{\cal I}}(t)$ 
is the same as (\ref{e9}) with 
$\mbf{D}(\mbf{r})=(e/(2m{\mathsf A}))
\hat{\mbf{z}}\times\mbf{r}$, 
corresponding to rotation. 
Using the WNA of Eq.(\ref{e12}), and dividing by 
$\omega^2$ as in (\ref{e34}) we get   
\begin{eqnarray} \label{e16} 
\tilde{C}_{{\cal I}}(\omega) = \left[
\frac{2}{3\pi}\frac{e^2}{{\mathsf A}^3} 
v_{\tbox{F}}^3 \oint |\mbf{n}\times\mbf{r}|^2 ds
\right] \frac{1}{\omega^2} 
\end{eqnarray}
Again, up to system-specific geometrical 
factor this result ($\sim \omega^{-2}$) agrees with the standard 
Drude result.
The latter expression should become exact 
as we go to large frequencies, where the 
only significant contribution comes form the 
self-correlation of the ${\cal F}(t)$ spikes (see Fig.~\ref{fig:f+c}d).

\begin{figure}
\centerline{\epsfig{figure=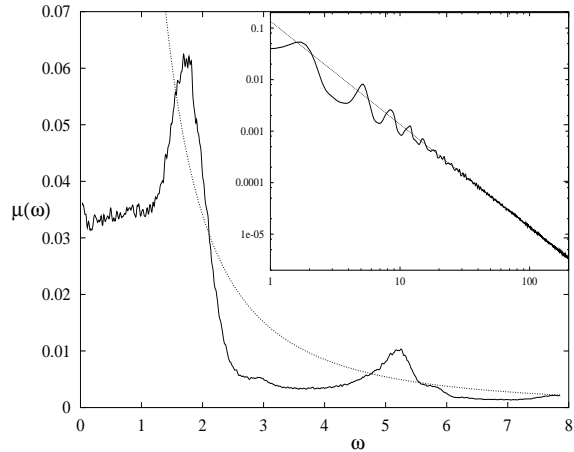,width=0.6\hsize}}
\caption{
Calculation of dissipation coefficient $\mu(\Omega)$ (arbitrary
units) for driving of a 
chaotic mesoscopic billiard system with a constant magnetic field at
frequency $\Omega$.
The billiard chosen is the 2D generalized Sinai of Fig.~\ref{fig:gsinai}a.
The dotted line WNA($\cal F$) is the high-frequency estimate
assuming ${\cal F}(t) \equiv \dot{{\cal I}}$ is white noise.
The convergence to this $\omega^{-2}$
result is clearly visible in the log-log inset plot.
}
\label{fig:drude}
\end{figure}

Eq.(\ref{e15}) leads  (via (\ref{e13}))  
to the small frequency Drude result, 
while the WNA of Eq.(\ref{e16}) gives  
the Lorentzian tail of the Drude result. 
An exact result for the frequency-dependent conductance 
can be calculated numerically for a given geometrical shape.
In Fig.~\ref{fig:drude} we display a plot of
$\mu(\Omega) \propto \tilde{C}_{{\cal I}}(\Omega)$, 
which shows both the constant behavior at small $\Omega$
and the convergence to the large-$\Omega$ WNA approximation.
System-specific features are expressed by the deviation  
from standard Lorentzian in the intermediate frequency regime.

Finally we consider
driving a quantum dot with homogeneous electric field in 
the $x$ direction,  in which case the Hamiltonian contains the 
interaction term $-e{\cal E}(t) x$. For calculation 
of the response in such a case one should evaluate the 
dipole-dipole correlation function $C_{\cal P}(\tau)$ 
where ${\cal P}(t) = ex$.  The latter is related 
to translations, where the deformation field 
is $\mbf{D} = \hat{\mbf{x}}$. Consequently 
we get $C_{\cal P}(\tau) = (1/\omega^4)C(\tau)$. 
However, this result is not of great interest, 
because the screening effect leads to modification 
of the effective one-particle Hamiltonian, such that  
the actual electric field inside a quantum dot 
is much smaller than the applied field.

\ \\

\section{The white noise assumption revisited}
\label{sec:wnarev}

In Section~\ref{sec:wna} we have assumed that
generic fluctuating quantities such as 
$\mbf{r}^2$ and $\mbf{e}\cdot\mbf{r}$ and 
$\mbf{e}\cdot(\mbf{r}\times\mbf{p})$  
have a white noise power spectrum
for $\omega \ll 1/\tbl$.
In section~\ref{sec:gen}
we are going to suggest that this white noise 
assumption is approximately true for any 
fluctuating quantity ${\cal F}(t)$ that comes 
from a normal deformation (the term `normal'  
will be defined there).

Obviously, the goodness of the `white noise assumption'
in the two cases mentioned
is related to the chaoticity of the system, and should be 
tested for particular examples.
This has been done so for the cavity of Fig.~\ref{fig:gsinai}
(see \cite{dil}, and Figs.~\ref{fig:wna_corr} and \ref{fig:add_goodgood}).
This cavity is an example of a `scattering billiard'
and so exhibits strong chaos \cite{bunim}.
If the motion is {\em not} strongly chaotic we may get a
$C(\tau)$ that decays like a power law (say $1/\tau^{1{-}\gamma}$ 
with $0<\gamma \le 1$) rather than an exponential
\cite{bunim,tails1,tails2,tails3,anomalous}.
In such case the universal behavior
is modified: we get $\omega^{-\gamma}$
behavior for $\cew$ at small frequencies ($\nu_{\tbox E}$ diverges),
signifying faster-than-diffusive energy spreading in Eq.(\ref{e25})
\cite{anomalous}.
The stadium is an example where such a complication 
may arise: an ergodic trajectory can remain in the
marginally-stable `bouncing ball' orbit
(between the top and bottom edges)
for long times, with a probability
scaling as $t^{-1}$ \cite{tails1,tails2,tails3}.
Depending on the choice of $D\ofs$ this
{\em may} manifest itself in $C(\tau)$.
For example, in Fig.~\ref{fig:qcc} the deformation P
involves a distortion confined to the upper edge,
and the resulting sensitivity to the bouncing ball orbit
leads to large enhancement of
the fluctuations intensity $\tilde{C}(\omega{=}0)$,
and is suggestive of singular behavior for small $\omega$.

If the billiard has a mixed phase space (which is the generic
case), then
the integrable component does not contribute to diffusive energy spreading.
Proposals have been made to account for this via a
phase-space volume factor \cite{mixed}.

\begin{figure}
\centerline{\epsfig{figure=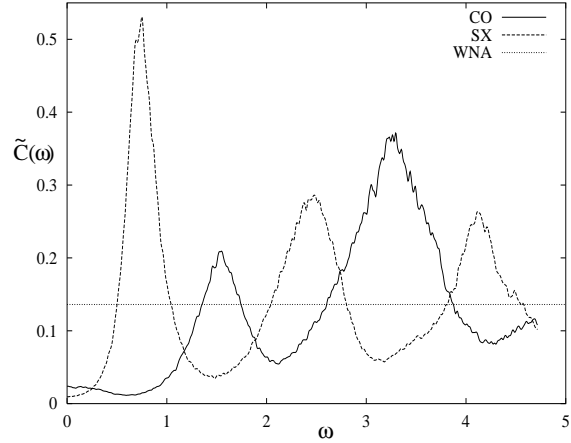,width=0.6\hsize}}
\caption{
The failure of the WNA estimate for $\tilde{C}(\omega)$
for deformation types CO (similar to DI)
and SX (similar to TX).
The WNA is clearly a vast overestimate of the small-$\omega$ limit.
See Tables~I and II for explanation of deformation types.
}
\label{fig:wna_gen}
\end{figure}
%

\section{Decomposition of general deformations}
\label{sec:gen}

The failure of the WNA for `special' deformations also
extends to the much wider class of deformations which are {\em similar}
to special. This is demonstrated in Fig. \ref{fig:wna_gen}.
It should be emphasized that this failure
happens even if the cavity is strongly chaotic.

We seek an analytical estimate for $\tilde{C}(\omega)$, 
and in particular for its zero-frequency limit $\nu$. 
This estimate should apply to any (general) deformation, 
including the case of `close-to-special' deformations. 
It would be useful to regard any general deformation  
as a combination of `special' component and `normal' component. 
The formulation of this idea is the 
theme of the present section.  Supporting numerical 
evidence is gathered in the next section.

The special deformations (for which we have $\nu=0$) 
constitute a linear space, meaning that any sum 
of special deformations is also a special one.
Now we would like to conjecture that there is also 
a linear space of `normal' deformations. By definition, 
for `normal' deformation ${\cal F}(t)$ looks like 
an uncorrelated random sequence of impulses, and 
consequently the WNA is a reasonable approximation.
The notion of randomness can be better formulated  
as in \ref{ap:cross} leading to Eq.(\ref{e37}). 
However in practice (\ref{e37}) is not useful, because 
it cannot be applied as an actual classification tool.  
(Eq.(\ref{e37}) is never satisfied exactly). 
Still we are going to demonstrate that there is a 
{\em unique} way to identify the subspace of normal deformations, 
if we insist on a maximal (\ie the most inclusive)
definition of this subspace.

It is important to clarify the heuristic reasoning of 
having linear space of normal deformations. 
The ${\cal F}(t)$ that corresponds to some normal 
deformation $D(\mbf{s})$ looks like white noise. 
It means that only self-correlations of its spikes 
are statistically significant. If we have 
two such generic quantities, 
say ${\cal F}_1(t)$  and ${\cal F}_2(t)$, 
then we expect ${\cal F}_1(t) + {\cal F}_2(t)$ 
to share the same property.

The correlation function 
of ${\cal F}(t) = {\cal F}_1(t) + {\cal F}_2(t)$ 
can be written formally as 
\begin{eqnarray} \label{e18} 
C_{1+2}(\tau) = C_1(\tau) + C_2(\tau) + 2C_{1,2}(\tau) 
\end{eqnarray}
where $C_{1,2}(\tau)$ is the {\em cross-correlation function}.
In \ref{ap:alex} we argue the following 
\begin{eqnarray} \nonumber 
\int_{-\infty}^{\infty} {C}_{1,2}(\tau) d\tau \ \ = \ \ 0 
\hspace*{3cm} \\ \label{e_19}
\mbox{if 1$=$general, 2$=$special}
\end{eqnarray}
This result is exact, and does not involve any approximation. 
In \ref{ap:cross} we argue the following 
\begin{eqnarray} \nonumber 
C_{1,2}(\tau) \; \approx \; c \times
	\left[
	\oint \! D_1(\mbf{s})D_2(\mbf{s}) d\mbf{s}
	\right]
	\delta(\tau)
\hspace*{0.5cm} \\ \label{e_20}
\ \ \ \ \ \mbox{if 1$=$normal, 2$=$general}
\end{eqnarray}
where 
$c=2m^2 v_{\tbox{E}}^3 \langle|\cos\theta|^3\rangle/{\mathsf V}$.
This result is an approximation, which is expected 
to be as good as our assumption regarding the 
`normality' of the deformation $D_1(\mbf{s})$.   
Consider now the case where $D_1(\mbf{s})$ is normal 
and $D_2(\mbf{s})$ is special. Both Eq.(\ref{e_19}) 
and Eq.(\ref{e_20}) should apply. But these equations 
are consistent if and only if $D_1(\mbf{s})$ is 
orthogonal to $D_2(\mbf{s})$. 
We say that $D_1(\mbf{s})$ and $D_2(\mbf{s})$ 
are orthogonal ($1\perp2$) using the 
following definition:    
\begin{eqnarray} \label{e20} 
\mbox{orthogonality} 
\ \ \ \ \ \Leftrightarrow \ \ \ \ \ 
\oint D_1(\mbf{s})D_2(\mbf{s}) d\mbf{s}  = 0
\end{eqnarray}
Thus we have proved that normal deformations must be 
orthogonal (in the sense of (\ref{e20})) to special 
deformations. Obviously we have proved here a necessary 
rather than a sufficient condition for `normality'. 
However, if we insist on a maximal definition for the 
subspace of normal deformations, then we get a unique 
identification. Namely, a deformation is classified 
as `normal' if it is orthogonal to the subspace of 
special deformations.

The practical consequences of Eq.(\ref{e_19}) 
and Eq.(\ref{e_20}) are as follows:
\begin{eqnarray} \label{e_21}
\nu_{1+2} = \nu_{1} 
\ \ \ \ \ \mbox{if 1$=$general, 2$=$special}
\end{eqnarray}
and 
\begin{eqnarray} \nonumber 
\nu_{1+2} \approx \nu_{1} + \nu_{2} + 
2c\oint \! D_1(\mbf{s})D_2(\mbf{s}) d\mbf{s}
\\ \label{e_22}
\ \ \ \ \ \mbox{if 1$=$normal, 2$=$general}
\end{eqnarray} 
These results are tested in the next section.

\section{Addition of deformations: numerical tests}
\label{sec:testnorm}

On the basis of the discussion in the previous section we {\em define} 
normal deformation as those that are orthogonal to all special deformations, 
in the sense of Eq.~(\ref{e20}). Obviously there are `good' normal 
deformations for which the WNA is an excellent approximation 
(P1 and W8 in Fig.~\ref{fig:wna_corr}, for example), and there are  
`bad' normal deformations for which the WNA
is not a very good approximation (FR in Fig.~\ref{fig:wna_corr},
and the normal component in Fig.~\ref{fig:decomp}b).
In this section we present numerical evidence that verifies
the theoretical results of the previous section,
and investigate how `bad' a normal deformation has to be for them
to break down.

\begin{figure}
\centerline{\epsfig{figure=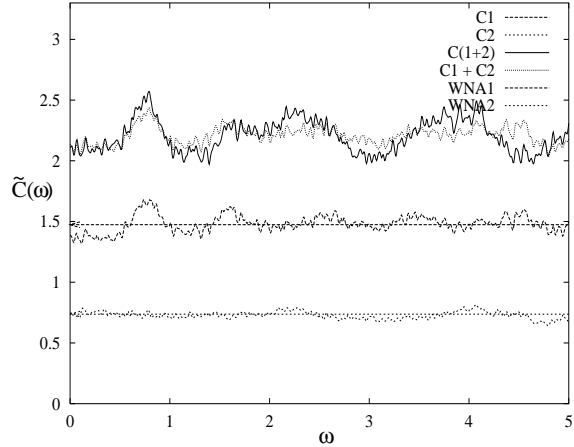,width=0.6\hsize}}
\caption{
Addition of two `good' normal deformations (1=P2, 2=WG).  
The two are orthogonal in the sense of (\ref{e20}). 
That they are `good' can be seen by their good agreement with their WNA
results (horizontal arrows).
The power spectrum of the sum agrees well with the sum of the power
spectra.
}
\label{fig:add_goodgood}
\end{figure}

From what we have claimed it follows that if
$D_1\ofs$ and $D_2\ofs$ are orthogonal 
normal deformations, then $\nu_{1+2}=\nu_1+\nu_2$.
We could as well write
\bea
	\tilde{C}_{1+2}(\omega) \; \approx \;
	\tilde{C}_1(\omega)+\tilde{C}_2(\omega)
\nonumber \hspace*{2cm} \\ 
\label{e21}
\ \ \ \ \ \mbox{if 1{=}normal, 2{=}normal, and 1 $\perp$ 2}
\eea
because by assumption the three correlation functions
are approximately flat.
We demonstrate this addition rule in
the case of two `good' deformations which are orthogonal in
Fig.~\ref{fig:add_goodgood}.
We found that small `pistons' (P2 is significant on only $\sim 1/50$ of the
perimeter) were needed to achieve addition of the
accuracy (a few \%) shown.
However, the restriction on the `wiggle' type of deformation
was somewhat more lenient (WG is $\sim 5$ times wider than P2 yet
obeys the WNA better than P2 does).

\begin{figure}
\centerline{\epsfig{figure=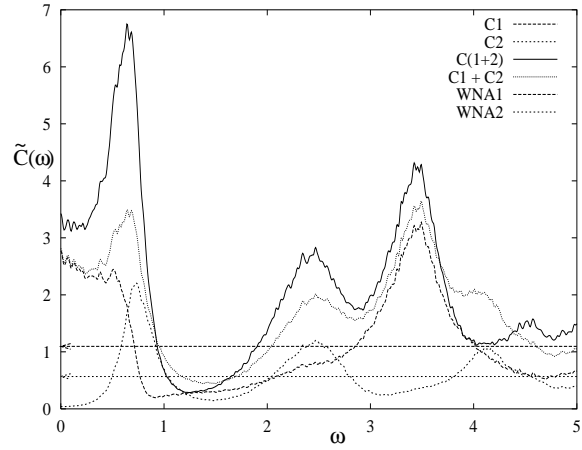,width=0.6\hsize}}
\caption{
Addition of two `bad' normal deformations (1=FR, 2=SX). 
The two are orthogonal in the sense of (\ref{e20}). 
That they are `bad' is shown by a lack of agreement with their WNAs.
The power spectrum of the sum is badly approximated 
by the sum of the power spectra (non-linear addition).
}
\label{fig:add_badbad}
\end{figure}

\begin{figure}
\centerline{\epsfig{figure=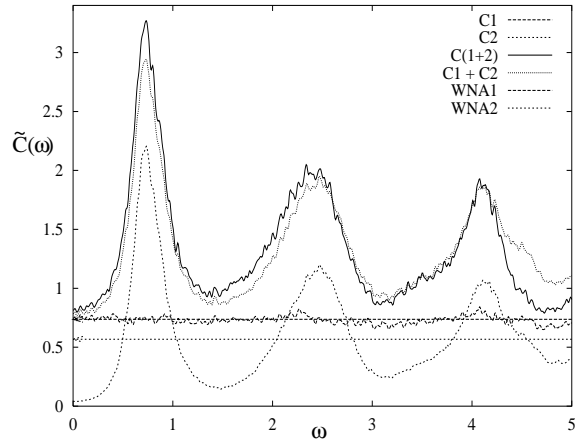,width=0.6\hsize}}
\caption{
Addition of a `good' normal deformation (1=WG) 
to a general deformation (2=SX).  
The two are orthogonal in the sense of (\ref{e20}).
The power spectrum of the sum agrees well with 
the sum of the power spectra. 
}
\label{fig:add_goodany}
\end{figure}

In general we observe that the quality of the addition
rule is limited by the
deviation from the WNA of the {\em better} of the two deformations.
In Fig.~\ref{fig:add_badbad} we see that 
if both $D_1(\mbf{s})$ and $D_2(\mbf{s})$ are bad, then also 
the addition rule (\ref{e21}) becomes quite bad.
Fig.~\ref{fig:add_goodany} shows that the addition rule (\ref{e21})
is reasonably well satisfied also 
if {\em either} $D_1(\mbf{s})$ or $D_2(\mbf{s})$ 
is a `good' normal deformation.
We have chosen $D_1(\mbf{s})$ as WG (good), 
and $D_2(\mbf{s})$ as SX which is almost completely
dominated by the special x-translation deformation.
The addition rule (\ref{e21}) is obeyed at all $\omega$. 
This proves that our assertions Eq.(\ref{e_20}) about the vanishing 
of $C^{\tbox{non-self}}_{1,2}(\tau)$ is indeed correct.  
It holds here as a non-trivial statement 
($D_2(\mbf{s})$ is general and `bad').

\begin{figure}
\centerline{\epsfig{figure=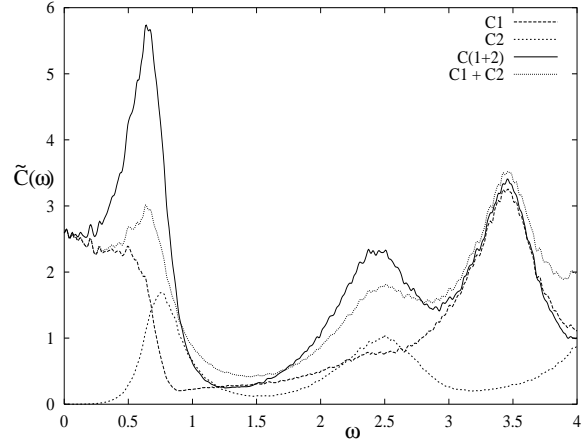,width=0.6\hsize}}
\caption{
Addition of a general deformation (1=FR) 
to `special' deformation (2=TX).
The power spectrum of the sum coincides with the 
sum of the power spectra in the limit $\omega\rightarrow0$, 
as implied by Eq.(\ref{e_21}). 
}
\label{fig:add_anyspec}
\end{figure}

Finally, we consider the case where $D_1(\mbf{s})$ is 
general and $D_2(\mbf{s})$ is special.
This is 
illustrated in Fig.~\ref{fig:add_anyspec}.
The addition rule (\ref{e21}) becomes exact in the limit 
of small frequency corresponding to the vanishing of
$\tilde{C}_{1,2}(\omega\rightarrow0)$ as implied by 
Eq.(\ref{e_19}). In particular this implies that $\nu_{1+2}=\nu_1$.
Note that there is {\em no} condition on the orthogonality of
$D_1(\mbf{s})$ and $D_2(\mbf{s})$).
This will be the key to for improving over the WNA, which 
we are going to discuss in the next section.

\begin{figure}
\centerline{\epsfig{figure=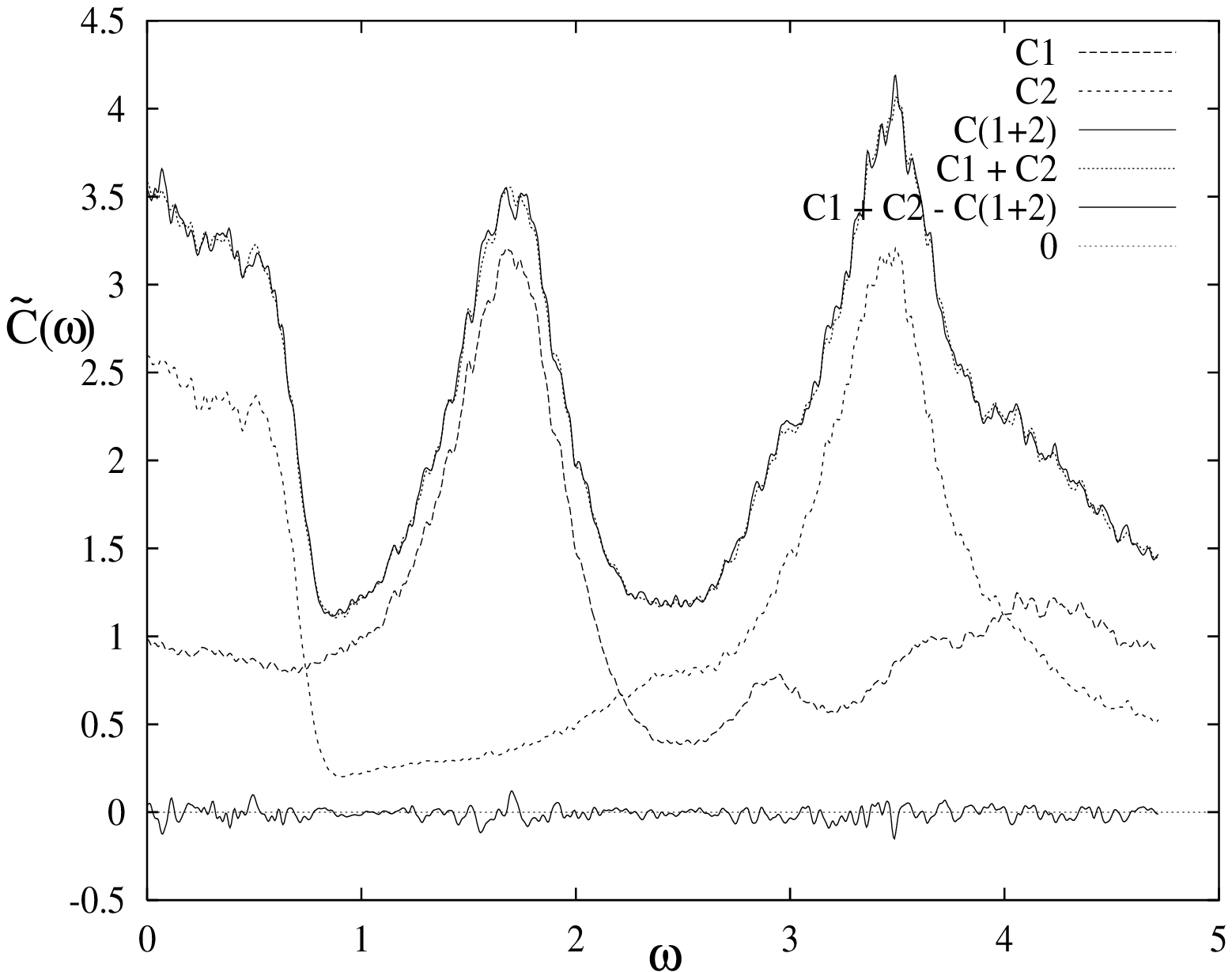,width=0.6\hsize}}
\caption{
Addition of two `bad' general deformations which come from different symmetry 
classes of the cavity (1=W2, 2=FR). 
The two must also be orthogonal, by symmetry.
The deviation from linear addition 
(solid line varying about zero) vanishes at all $\omega$.
}
\label{fig:add_symm}
\end{figure}

In drawing the above conclusions it is important to note 
that symmetry effects can play a deceptive role
if the cavity shape has symmetry
(our example Fig.~\ref{fig:gsinai} is in the $C_{2v}$ symmetry group).
In Fig.~\ref{fig:add_symm} we demonstrate that the addition rule (\ref{e21})
is very accurately satisfied at {\em all} $\omega$ 
if $D_1(\mbf{s})$ and $D_2(\mbf{s})$
belong to different symmetry classes of the cavity.
Orthogonality of $D_1(\mbf{s})$ and $D_2(\mbf{s})$
is {\em not} sufficient to explain
this perfect linearity of addition of $\cew$.
Rather, it follows from the symmetry of the kernel
$\gamma_{\tbox E}(\mbf{s}_1,\mbf{s}_2)$ of Eq.(\ref{eq:kernel}). 
The cross-terms in (\ref{eq:kernel}) rigorously vanish
when such deformations are added.
The consequence is that in order to demonstrate the
assertions of this and of the previous section, 
we had to choose deformations of the {\em same}
symmetry class, or which break all symmetries of the cavity.

%

\section{Beyond the WNA}
\label{sec:beywna}

\begin{figure}
\centerline{\epsfig{figure=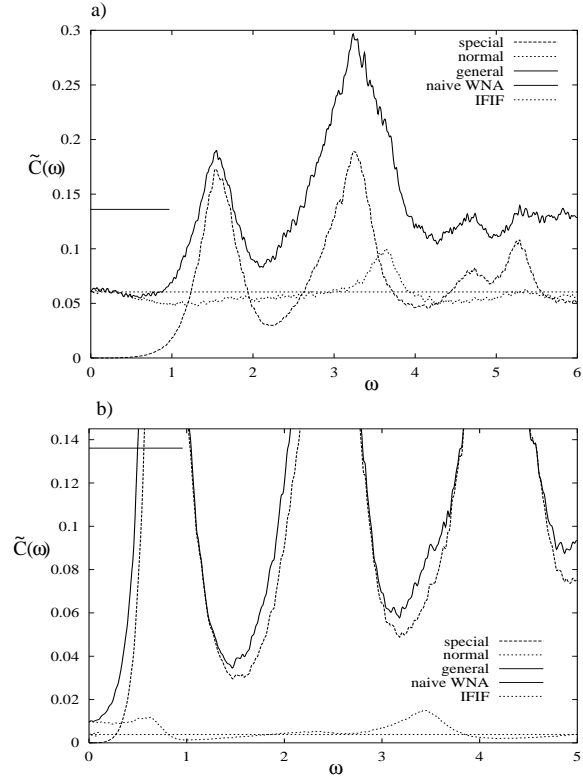,width=0.6\hsize}}
\caption{
Decomposition of general deformations $D(s)$ into
orthogonal `normal' and `special' components. 
The general deformation is CO + W16 in subfigure (a), 
and SX in subfigure (b). 
The naive WNA Eq.(\ref{e12}) is indicated by short solid line. 
The improved (IFIF) result Eq.(\ref{e23}) is indicated by long dashed arrow.
}
\label{fig:decomp}
\end{figure}

It is possible now to consider the case of general 
deformation, and to go beyond the WNA.
Given a general 
deformation $D(\mbf{s})$ we should project out (subtract) all the special 
components, leaving the normal component,
and only then apply the WNA.
In Fig.~\ref{fig:decomp} we demonstrate 
this decomposition for the deformation
(CO + W16) and the deformation SX.

The special deformations constitute a linear space
which is spanned by the basis functions: one dilation, $d$ translations, 
and $d(d-1)/2$ rotations.
(For $d{=}2$ they are listed in Table~II).
For a general cavity shape these basis functions are not orthogonal.
However, because they are linearly independent,
we can use standard linear algebra to build an orthonormal
basis $\{ D_i(\mbf{s}) \}$ of special deformations.
The special ($\parallel$) and the normal ($\perp$) components of any given
deformation $D(\mbf{s})$ are therefore
\begin{eqnarray} \nonumber
D_{\parallel}(\mbf{s}) &=& 
\mbox{$\sum_i \alpha_i D_i(\mbf{s})$} 
\\ \label{e_dcmp}
D_{\perp}(\mbf{s}) &=& D(\mbf{s})-D_{\parallel}(\mbf{s})
\end{eqnarray} 
where the coefficients are
\be
\label{eq:alphai}
	\alpha_i \; = \; \oint \! D(\mbf{s}) D_i(\mbf{s}) \, d\mbf{s}.
\ee
The improved approximation for $\nu$ applies the WNA
only to the normal component, giving
\begin{eqnarray} \label{e23}  
	\nu_{\tbox{E}} \ \approx \ 
	2m^2 v_{\tbox{E}}^3\langle|\cos\theta|^3\rangle 
	\frac{1}{{\mathsf V}} 
	\oint d\mbf{s} [D_{\perp}(\mbf{s})]^2
\end{eqnarray}
which we name the IFIF (Improved Fluctuations Intensity Formula).
In the particular case of $d{=}3$,
substitution of this result into the microcanonical FD relation
gives an `improved wall formula'
consisting of the replacement of
$D\ofs$ by $D_{\perp}\ofs$ in Eq.~(\ref{e4}).

In Fig.~\ref{fig:decomp}
we use the IFIF to estimate $\nu$ for two examples.
The first is a deformation (CO + W16) whose normal component
is `good', due its oscillatory nature. The deviation 
from a flat white power spectrum is $\sim 20$\% for the 
normal component. 
The IFIF result Eq.(\ref{e23}) is accurate to a few percent. 
It is a much better estimate of the actual $\nu$ compared 
with the naive WNA Eq.(\ref{e12}) which overestimates 
the correct value by a factor of 2.2.
In the second example the deformation is SX. 
The resulting normal component is `bad'. Its 
power spectrum fluctuates by a factor of about 10
in the $\omega$ range shown.
Consequently the IFIF
is limited in its accuracy, and the correct value 
for $\nu$ is underestimated by a factor of 2.5.
However, it is still a great improvement over the 
naive result Eq.(\ref{e12}).
In this second example we can extract another prediction 
about $\cew$. The special component is a factor $\sim 10$ larger
than the normal component. Therefore 
the $\omega^2$ behavior at small $\omega$ is almost entirely 
due to the `rotation' component. The prefactor of the 
$\omega^2$ behavior need only be found once for each billiard shape
(see Section~\ref{sec:drude}). This saves computation and 
gives an extra information about the dissipation rate 
at finite driving frequency.

\begin{figure}
\centerline{\epsfig{figure=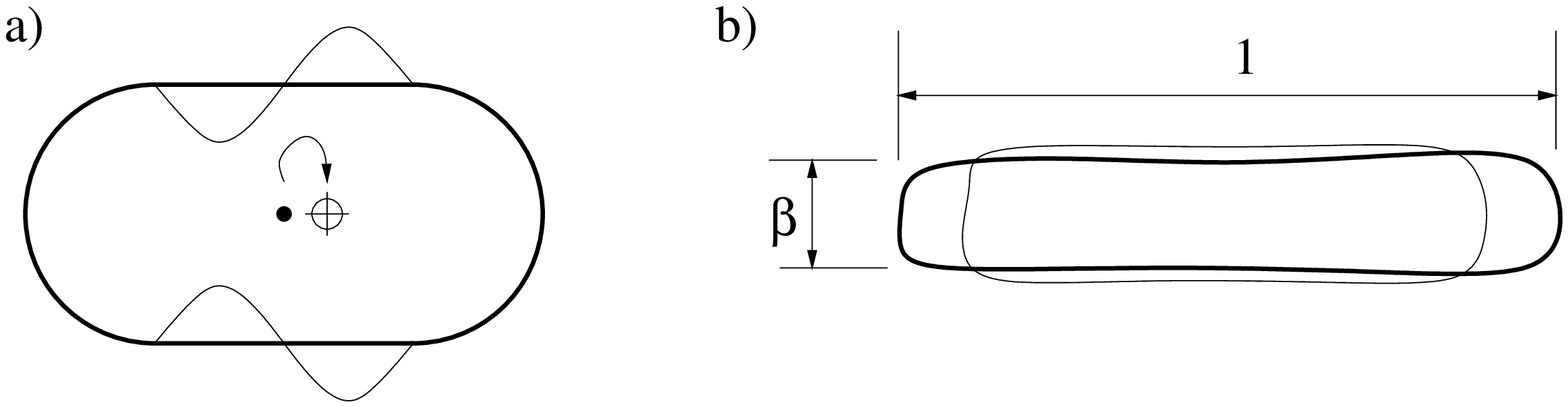,width=0.7\hsize}}
\caption{
{\bf a}) A deformation of the stadium which moves the
`center of mass' (centroid) of the cavity to the right
(from the dot to the crosshairs symbol).
This deformation is orthogonal
(in the sense of (\ref{e20}))
to all special deformations,
in particular, all translations.
{\bf b})~An example volume-preserving deformation of an elongated
approximately-rectangular
cavity ($\beta \ll 1$) which nevertheless has a large overlap
with dilation.
It can be shown that this results in an IFIF
estimate of $\approx 4\beta$ times
that of the naive WNA.
In both diagrams the undeformed shape is shown as a heavy line,
the deformed one as a thin line.
}
\label{fig:history}
\end{figure}

A few concluding remarks regarding the history of  
the wall formula are in order.
It has been known since its inception that the naive wall formula gives 
unphysical answers in the case of {\em constant-velocity}
translations and rotations. This was first regarded  
as a kinetic gas `drift' effect \cite{wall}.
It should be noted that the recipe presented in \cite{wall}, 
namely to subtract this drift component,
is equivalent in practice to the recipe (\ref{e_dcmp}) that we have 
presented here, {\em provided we ignore dilations}. 
It is also important to realize that the 
argumentation in \cite{wall} for this subtraction 
appears to be ad hoc, being based on a
`least-structured drift pattern' reasoning.
A stated condition on this subtraction
was that the resulting deformation preserve the location
of the `center of mass' (centroid) of the cavity,
for reasons particular to the nuclear application \cite{wall}. 
This condition seems to have become standard practice in
numerical tests of the wall formula \cite{mixed,test1,test2,test3}.
However, as the system in Fig.~\ref{fig:history}a illustrates,
this condition is generally {\em not} equivalent to the above subtraction
of translation and rotation components \cite{cofm}.
This seems to invalidate the theorem presented in Section 7.1 of \cite{wall}.
Where the flaw in their reasoning lies we are not sure.

The consideration of the special nature of
dilations is absent from the literature.
Even if we restrict ourselves to volume-preserving deformations
(the case for the nuclear application),
then deformations of certain cavities can be found for which
the dilation correction is significant; we illustrate this in
Fig.~\ref{fig:history}b.
This correction can only be large if the cavity has a large
variation in radius (\ie is highly non-spherical).
We suggest this as a possible reason why major discrepancies
due to dilation
have not emerged
in the
numerical tests of the wall formula until now.
Such tests have generally been of shapes close to a 3D sphere
\cite{wall,mixed,test1,test2,test3}.

Hence we believe that the recipe we have presented, 
along with the associated theory and in conjunction with 
the particular power-law dependences, is a significant 
step in the treatment of one-body dissipation.

\ack{
In particular we would like to thank Eduardo Vergini
for stimulating dialogue which motivated us to consider
the special nature of dilations.
This work was funded by ITAMP
(at the Harvard-Smithsonian Center for Astrophysics
and the Harvard Physics Department)
and the National Science Foundation (USA).
}

\newpage

\appendix
\section{Linear response theory of dissipation}
\label{ap:lrt}

Given a parametric Hamiltonian $H(Q,P;x(t))$, and given initial 
conditions, one defines the energy 
${\cal E}(t)=H(Q(t),P(t);x(t))$ and the fluctuating quantity 
${\cal F}(t)=-{\partial {\cal H}}/{\partial x}(Q(t),P(t);x(t))$. 
With no approximation we have 
\begin{eqnarray} \label{e24}
{\cal E}(t)-{\cal E}(0) = \int_0^t {\cal F}(t') \dot{x}(t') dt' .
\end{eqnarray} 
Using the same steps as in \cite{frc} one obtains 
the following result for the variance of the energy spreading:
\begin{eqnarray} \label{e25}
\delta E(t)^2 = \int_0^t\int_0^t  C_{\tbox E}(t''-t') F(t''-t') dt' dt''  
\end{eqnarray} 
where $C_{\tbox E}(\tau)$ is defined by Eq.(\ref{e8}).
Microcanonical averaging has been taken over the 
initial conditions. The function 
$F(\tau)=\langle \dot{x}(t)\dot{x}(t+\tau) \rangle$ 
is the velocity-velocity correlation of the driving.  
For periodic driving $x(t) = A\sin(\Omega t + \mbox{\small phase})$ 
it is formally convenient to average over the initial 
phase and one obtains $F(\tau)=\half (A\Omega)^2 \cos(\Omega\tau)$.

For a chaotic system $C_{\tbox E}(\tau)$
is characterized by some correlation 
time $\tau_{\tbox{cl}}$.
For $t\gg\tau_{\tbox{cl}}$
one obtains diffusive spreading 
$\delta E(t)^2 = 2D_{\tbox{E}} t$ where the diffusion rate is
\begin{eqnarray} \label{e26}
D_{\tbox{E}} =
\frac{1}{2} \tilde{C}_{\tbox E}(\Omega) \times \half (A\Omega)^2 
\end{eqnarray} 
which for small frequencies goes to
$D_{\tbox{E}} = \half \nu_{\tbox{E}} V^2$,
where $\nu_{\tbox{E}} \equiv \tilde{C}(0)$ as defined in the Introduction.
The picture to keep in mind is that of
the fluctuating ${\cal F}(t)$ causing a {\em random walk}
in energy space via Eq.(\ref{e24}) for times $t\gg\tau_{\tbox{cl}}$.
As explained in \cite{jarz92,jarz93,frc} the
resulting diffusion in energy space
implies systematic {\em growth} of the average energy.
It is important to realise that this growth happens
even if the random walk is locally unbiased:
such is the case when changing the parameter $x$ preserves
the volume of a given energy-shell in phase-space.
(For a deforming billiard system this corresponds to 
preservation of the billiard volume).
The rate of 
energy growth is related to the diffusion as follows:
\begin{eqnarray} \label{e27}
\frac{d}{dt}\langle {\cal H} \rangle 
\ = \ - \int_0^{\infty} dE \ g(E) \ D_{\tbox{E}}  
\ \frac{\partial}{\partial E}
\left(\frac{\rho(E)}{g(E)}\right)
\end{eqnarray} 
where $\rho(E)$ is the energy distribution of the particles, 
and $g(E)$ is the one-particle density of states.
The growth is therefore an effect of the $E$-dependence of both
the diffusion rate and the density of states.

The rate of dissipation can be written as in Eq.(\ref{e1a}) 
or as $d\langle{\cal H}\rangle/dt =  \mu V^2$ 
in the small frequency limit. 
Combining this with Eq.(\ref{e27})
implies a relation between the dissipation 
coefficient $\mu$ and the function $\tilde{C}_{\tbox E}(\omega)$.
The most familiar version of this FD relation is obtained 
for small frequency under the assumption of a
canonical distribution $\rho(E) \propto g(E) \exp(-E/(k_BT))$, 
leading to 
\begin{eqnarray} \label{e28}
\mu = \frac{1}{2k_BT} \nu 
\end{eqnarray} 
where $\nu$ should be calculated for a canonical distribution.  
This result should be multiplied by the number of non-interacting 
classical particles.

The use of Eq.(\ref{e27}) can be justified also for non-interacting 
fermions \cite{wilk90}.
This is because the effect of Pauli exclusion principle 
cancels out (in analogy with Boltzmann picture with elastic scattering).
Substituting $\rho(E)=g(E)f(E-E_{\tbox{F}})$, where 
$f(E-E_{\tbox{F}})$ is the Fermi occupation function, one obtains
\begin{eqnarray} \label{e29}
\mu = \frac{1}{2} g(E_{\tbox{F}}) \nu_{\tbox{F}} 
\end{eqnarray} 
where $\nu_{\tbox{F}}$ should be calculated at the Fermi energy.

Finally, the microcanonical version of the FD relation is 
\begin{eqnarray} \label{e30}
\mu_{\tbox{E}} \ = \ \frac{1}{2} \frac{1}{g(E)} 
\frac{\partial}{\partial E}
(g(E) \nu_{\tbox{E}}) 
\end{eqnarray} 
The subscript E indicates that both $\nu_{\tbox{E}}$ and $\mu_{\tbox{E}}$ 
are evaluated locally around some energy $E$.

\section{Cross correlations I}
\label{ap:alex}

In this Appendix we introduce two proofs of Eq.(\ref{e_19}). 
The first is a formal argument, while the second 
is a more physically appealing argument.  
The formal argument is as follows: 
Eq.(\ref{eq:kernel}) is an exact result which can be written 
using obvious abstract matrix notation as 
$\nu = D \gamma_{\tbox E} D$.  Let  $D=D_1+D_2$. 
If $D_2$ is a special deformation then 
by definition $D_2 \gamma_{\tbox E} D_2  = 0$.
But this can be true only if  $D_2$ belongs 
to the kernel (nullspace) of the matrix $\gamma_{\tbox E}$, 
hence we have  $\gamma_{\tbox E} D_2  = 0$. 
Therefore we have also $D_1 \gamma_{\tbox E} D_2  = 0$ 
for any  $D_1$, which is precisely the statement 
of Eq.(\ref{e_19}).

Now we present the alternative physically appealing argument.
Consider two noisy signals ${\cal F}(t)$ and ${\cal G}(t)$.
We assume that
$\langle {\cal F}(t)\rangle  = \langle {\cal G}(t)\rangle = 0$.  
The angular brackets stand for an average over realizations. 
The auto-correlations of ${\cal F}(t)$ and ${\cal G}(t)$ 
are described by functions $C_{\tbox{F}}(\tau)$ and 
$C_{\tbox{G}}(\tau)$ respectively. We assume that both 
auto-correlation functions are short-range, meaning 
no power-law tails (this corresponds to the hard chaos 
assumption of this paper),
and that they are negligible beyond a time $\tau_{\tbox c}$.
We call a signal `special' if the algebraic area under its auto-correlation
is zero.
The cross-correlation function is defined as 
\begin{eqnarray} \label{e_1} 
	C_{\tbox{F,G}}(\tau) \equiv
	\langle {\cal F}(t') {\cal G}(t'') \rangle ,
	\hspace{.5in}
	\tau \equiv t'-t'' .
\end{eqnarray} 
We assume stationary processes so that the cross-correlation 
function depends only on the time difference $\tau$. 
We also symmetrize this function if it does not have 
$\tau\mapsto-\tau$ symmetry. We assume that $C_{\tbox{F,G}}(\tau)$ 
is short-range, meaning that it becomes negligibly small  
for  $|\tau| > \tau_{\tbox{c}}$.
We would like to prove that if either ${\cal F}(t)$  or 
${\cal G}(t)$ is special
then the algebraic area under the cross-correlation 
function equals zero.

Consider the case where ${\cal F}(t)$ is general
while ${\cal G}(t)$ is special.
The integral of $C_{\tbox{F}}(\tau)$ will be denoted by $\nu$.
Define the processes 
\begin{eqnarray}
\label{e_2a} 
	X(t) &=& \int_0^t {\cal F}(t') dt'  \\
\label{e_2b}
	Y(t) &=& \int_0^t {\cal G}(t'') dt'' .
\end{eqnarray} 
From our assumptions it follows, disregarding 
a transient, that for $t \gg \tau_{\tbox c}$ we have
diffusive growth $\langle X(t)^2 \rangle \approx \nu t$.
However since $Y(t)$ is a stationary process \cite{gardiner},
$\langle Y(t)^2 \rangle \approx \mbox{const}$.
Therefore for a typical realization we have 
$|X(t)| \le \mbox{const} \times \sqrt{\nu t}$
and $|Y(t)| \le \mbox{const}$. Consequently, without 
making any claims on the independence of $X(t)$ and $Y(t)$,
we get that $\langle X(t) Y(t) \rangle$ cannot grow
faster than $\mbox{const} \times \sqrt{ \nu t}$.
Using the definitions (\ref{e_2a}), (\ref{e_2b})
and (\ref{e_1}) we can write
\bea
\label{e_3}
	\int_{-\infty}^{\infty} C_{\tbox{F,G}}(\tau) d\tau
	=
	\frac{\langle X(t) Y(t) \rangle}{t}
	\approx
	\frac{\mbox{const}}{\sqrt{t}} \rightarrow 0
\eea
where the limit $t\rightarrow\infty$ is taken.
Thus we have proved our assertion.

\section{Cross correlations II}
\label{ap:cross}

In this section we further discuss some features 
of the cross-correlation function. 
For the purpose of presentation we 
we would like to view the time as an integer 
variable $t=1,2,3...$. One may think of each instant
of time as corresponding to a bounce.

Let us assume that we have functions $f(s)$ and $g(s)$,  
and a time-sequence $(s_1,s_2, s_3,...)$.
This gives two stochastic-like processes 
$({\cal F}_1, {\cal F}_2, {\cal F}_3,...)$ and 
$({\cal G}_1, {\cal G}_2, {\cal G}_3,...)$.
The cross correlation of these two processes is 
defined as follows: 
\begin{eqnarray} \label{e35} 
C_{\tbox{F,G}}(i-j) = \langle  {\cal F}_i  {\cal G}_j  \rangle 
= \langle f(s_i) g(s_j) \rangle   
\end{eqnarray} 
It is implicit in this definition that we assume 
that the processes are stationary, so the result depends 
only on the difference $\tau=(i-j)$. 
The angular brackets stand for an average over realizations 
of $s$-sequences.

If the sequences are ergodic on the $s$ domain, 
then it follows that 
\begin{eqnarray} \nonumber
\langle {\cal F} \rangle &=& \int \! f(s) ds 
\\ \nonumber
\langle {\cal G} \rangle &=& \int \! g(s) ds 
\\
\label{eq:ergself}
	C_{\tbox{F,G}}(0) &=& \int \! f(s)g(s) ds 
\end{eqnarray} 
The $\tau \ne 0$ cross-correlations requires information 
beyond mere ergodicity. In case that the $s$ sequence 
is completely uncorrelated in time we can factorized the 
averaging and we get 
$C_{\tbox{F,G}}(\tau \ne 0) = 
\langle {\cal F} \rangle  \times 
\langle {\cal G} \rangle$. 
If $\langle {\cal F} \rangle = 0$ then 
\begin{eqnarray} \label{e36} 
	C_{\tbox{F,G}}(\tau \ne 0) = 0
\end{eqnarray} 
irrespective of $\langle {\cal G} \rangle$.

However, we would like to define circumstances 
in which Eq.(\ref{e36}) is valid, even if 
the $s$ sequence is {\em not} uncorrelated. 
In such case either the ${\cal F}$ or the 
${\cal G}$ may possess time correlations. 
(Such is the case if ${\cal G}$ is `special'). 
So let us consider the case where the ${\cal F}$ 
sequence {\em looks random}, while assuming nothing about 
the ${\cal G}$ sequence. By the phrase `looks random' 
we mean that the conditional probability satisfies 
\begin{eqnarray} \label{e37} 
\mbox{Prob}({\cal F}_i | s_j) = \mbox{Prob}({\cal F}_i)
\ \ \ \ \ \ \mbox{for any $i \ne j$}
\end{eqnarray} 
Eq. (\ref{e36}) straightforwardly follows 
provided $\langle {\cal F} \rangle = 0$, 
irrespective of the $g(s)$ involved. 
Given $f(s)$, the goodness of assumption (\ref{e37}) 
can be actually tested. However, it is not 
convenient to consider  (\ref{e37}) as a practical  
definition of a `normal' deformation.

\newpage
      

\end{document}